\documentstyle[aps,preprint]{revtex}
\input{psfig}
\begin{document}
\title{\bf Field purification in the intensity-dependent Jaynes-Cummings model}
\author{Dagoberto S. Freitas,\footnote{dfreitas@ifi.unicamp.br}
\thanks{Permanent address:
Universidade Estadual de Feira de Santana, Departamento de Ci\^encias Exatas, 
44031-460, Feira de Santana, BA, Brazil}
{\/\/}A. Vidiella-Barranco\footnote{vidiella@ifi.unicamp.br}, 
and J.A. Roversi\footnote{roversi@ifi.unicamp.br}}
\address{Instituto de F\'\i sica ``Gleb Wataghin'',
Universidade Estadual de Campinas,
13083-970   Campinas  SP  Brazil.}
\maketitle
\begin{abstract}
We have found that, in the intensity-dependent Jaynes-Cummings model, 
a field initially prepared in a statistical mixture of two coherent states, 
$|\alpha\rangle$ and $|-\alpha\rangle$,
evolves toward a pure state.
We have also shown that an even-coherent state turns
 
periodically a into rotated odd-coherent state during the evolution.
\end{abstract}
\pacs{42.50.Dv, 42.50.Ct} 


\section{Introduction}

The generation of nonclassical light and its interaction with matter are subjects
of intense investigation in quantum optics. The increasing control of atoms and
electromagnetic
fields achieved nowadays has opened up exciting possibilities in this field. 
Details of the matter-field interaction have been investigated over the past thirty 
years, specially after the introduction of the Jaynes-Cummings model \cite{jayn}. 
Despite of the simplicity of the model, it allows generalizations that may be applied
to different circumstances and regimes \cite{knig}. One of these generalizations 
is the intensity dependent Jaynes-Cummings model, introduced by Buck and Sukumar 
\cite{buck}. Because of the commensurability of the Rabi frequencies which 
arises from such a coupling, this model presents absolutely periodic revivals, 
contrary to what happens in the ordinary Jaynes-Cummings model. 
Moreover, the state-vector representing the evolution
of the system is periodic itself. This means that there will be periodic evolution for
any expectation value. What has not been acknowledged is that this behaviour 
leads to such an {\it enhancement of certain effects\/} that would be otherwise 
difficult to notice within the realm of the original Jaynes-Cummings model. 
Because of this enhancement it is possible to have the generation of well-defined
Schr\"odinger cat-like states during the evolution of the field in the 
intensity-dependent model, as it has been already discussed \cite{zahe}. We
would like to remark that the approach to a (almost) pure state at
half of the revival time occurs in the ordinary Jaynes-Cummings model, as
it is well known \cite{phoe,geaba}, if we start with the field in a pure state.
For an initial statistical mixture, however, only a tendency of 
purification occurs, instead of perfect purification.

In this paper we are going to be concerned with the dynamical change 
of field states, namely superpositions of coherent states. Of particular 
importance is the purification, i.e., the  
transformation of a statistical mixture of two coherent states, for instance, 
into a quantum superposition of coherent states. We know that normally processes 
such as the interaction of a field with its environment (leading to dissipation), 
and the resonant interaction of a field with atoms leads to important loss of
coherence, which represents the destruction of the quantum properties of a 
field state. It is therefore important to look for ways of overcoming these 
very common demolition processes. Surprisingly enough we have found that the 
resonant intensity-dependent Jaynes Cummings model provides a possible purification 
procedure. This is connected to the intrinsic periodicity of the model, and it is
easy to see how this reorganization occurs from the phase-space point of view. 
On the other hand, if we start with the field prepared in a (pure) 
even-coherent state, the model shows periodic revivals which occur at half of the
time of the revivals for an initial coherent state \cite{vidi}.  
Because the full periodicity of the model again, the atom returns to its initial 
state (the excited state, for instance) at the second revival. However, at the 
first revival the atom will invert its state, i.e., it will appear in the ground
state. The atom-field disentanglement (with relatively high intensity fields) 
at that time guarantees that the field will be in a pure state, but due to the change 
in the atomic state, we expect that not to be the even-coherent state. 
In fact, the field at that first revival time becomes a rotated odd-coherent state, 
as we are going to show. 

This paper is organized as follows: in Section II we present the solution of the model
in terms of the density operator. We also
show the evolution of the field purity. In Section III we discuss the purification 
procedure from the phase space point of view. In Section IV we analyze the evolution 
of fields initially prepared in pure states (even and odd coherent states). 

\section{Density operator solution and the field purity}

The Hamiltonian for the intensity-dependent Jaynes-Cummings model \cite{buck} under 
the RWA is given by  

\begin{equation}
\hat{H}=\frac{\hbar \omega}{2}\sigma _{z}+\hbar \omega\left(\hat{a}^{\dagger}
\hat{a}+\frac{1}{2}\right)+
\hbar\lambda\left(\hat{R}\sigma _{+}+\sigma _{-}\hat{R}^{\dagger }\right),
\label{eq:hamil}
\end{equation}
where $\lambda$ is the usual atom-field coupling constant, $\sigma_{+}=\left|
e\right\rangle \left\langle g\right| $ and $\sigma_{-}=\left|
g\right\rangle \left\langle e\right| $ are the atomic creation and
annihilation operators respectively and $\hat{R}=\hat{a}\left( 
\hat{a}^{\dagger }\hat{a}\right) ^{1/2}$, $\hat{R}^{\dagger
}=\left( \hat{a}^{\dagger }\hat{a}\right) ^{1/2}\hat{a}^{\dagger
}$. Because of the factor $\left( \hat{a}^{\dagger }\hat{a}\right) ^{1/2}$, 
the interaction term is no longer linear in the field variables and represents
an intensity-dependent coupling. Let us assume that the initial state of the
system is the product state $\rho _{af}\left( 0\right) =\rho _{a}\left(
0\right) \otimes \rho _{f}\left( 0\right) $, with the atom initially in the
upper state, or $\rho _{a}\left(0\right)=|e\rangle\langle e|$. The solution for the
time-dependent density operator is analogous to the one in the ordinary 
Jaynes-Cummings model \cite{phoe,sten}, so that the evolution operator in the 
(two-state) atomic basis is
\begin{equation}
U^{\dagger }(t)=\left[ 
\begin{array}{cc}
\hat{C}_{n+1}(t) & i\hat{S}_{n+1}(t)\hat{R} \\ 
i\hat{S}_{n}(t)\hat{R}^{\dagger } & \hat{C}_{n}(t)
\end{array}
\right] ,  \label{eq:moevt}
\end{equation}
where

\begin{equation}
\hat{C}_{n+1}=\cos\left(\lambda t\sqrt{\hat{R}\hat{R}^{\dagger }}\right),\ \ \ \ \ \
\hat{S}_{n+1}=\frac{\sin \left(\lambda t\sqrt{\hat{R}
\hat{R}^{\dagger}}\right)}{\sqrt{\hat{R}\hat{R}^{\dagger}}},
\label{eq:emopd}
\end{equation}
and having the same expressions for $\hat{C}_{n}$ and $\hat{S}_{n}$ but with 
$\hat{R}^{\dagger}\hat{R}$ instead of $\hat{R}\hat{R}^{\dagger}$.

Therefore the time-evolved density operator will read
\begin{equation}
\hat{\rho}_{af}(t)=U(t)\hat{\rho}_{af}(0)U^{\dagger }(t)=
\left[ 
\begin{array}{cc}
\hat{A}(t)\hat{\rho}_f(0)\hat{A}^{\dagger }(t) & \hat{A}(t)\hat{\rho}_f(0)
\hat{B}^{\dagger }(t) \\ 
\hat{B}(t)\hat{\rho}_f(0)\hat{A}^{\dagger }(t) & \hat{B}(t)\hat{\rho}_f(0)
\hat{B}^{\dagger}(t)
\end{array}
\right],
\end{equation}
where $\hat{A}(t)=\hat{C}_{n+1}$ and 
$\hat{B}(t)=-i\hat{R}^{\dagger }\hat{S}_{n+1}$. After tracing over the atomic 
variables we obtain the reduced field density operator, or 
$\hat{\rho}_f(t)=Tr_a\left[\hat{\rho}_{af}(t)\right]$, which is given by
\begin{equation}
\hat{\rho }_{f}(t)=\hat{A}(t)\hat{\rho }_{f}(0)\hat{A}%
^{\dagger }(t)+\hat{B}(t)\hat{\rho }_{f}(0)\hat{B}^{\dagger }(t).
\label{eq:odrf}
\end{equation}
If the state of the initial field is an equally-weighted statistical mixture
of two coherent states, or 
\begin{equation}
\hat{\rho}_{f}(0)=\frac{1}{2}\left( |\alpha \rangle \langle \alpha
|+|-\alpha \rangle \langle -\alpha |\right),\label{eq:insta}
\end{equation}
we have that
\begin{eqnarray}
\hat{\rho}_f(t)&=&\frac{1}{2}\sum_{n,m}\frac{e^{-|\alpha|^2}}{\sqrt{m!}\sqrt{n!}}
\alpha^n\alpha^*{}^m\left[1+(-1)^{n+m}\right]\Big\{\cos\left[\lambda t(n+1)\right]
\cos\left[\lambda t(m+1)\right]|n\rangle\langle m|\nonumber \\
&+&\frac{\sqrt{n+1}}{\alpha}
\sin\left[\lambda t(n+1)\right]\sin\left[\lambda t(m+1)\right]
|n+1\rangle\langle m+1|\Big\}\label{eq:odint}.
\end{eqnarray}
For the sake of simplicity we are going to consider the amplitude $\alpha$ as real.
At half of the revival time ($t=\pi/2\lambda$), we note that, for 
$\overline{n}\gg 1$, each one of the terms in Eq. (\ref{eq:odint}) become exactly
equal to
\begin{equation}
\hat{A}(\pi/2\lambda)\hat{\rho }_{f}(0)\hat{A}^{\dagger }(\pi/2\lambda)=
\hat{B}(\pi/2\lambda)\hat{\rho }_{f}(0)\hat{B}^{\dagger }(\pi/2\lambda)=
\frac{1}{4}\left(|i\alpha\rangle-|-i\alpha\rangle\right)
\left(\langle i\alpha|-\langle -i\alpha|\right),
\end{equation}
so that he resulting field state will be equal to
\begin{equation}
\hat{\rho}_f(\pi/2\lambda)=\frac{1}{2}\left(|i\alpha\rangle-|-i\alpha\rangle\right)
\left(\langle i\alpha|-\langle -i\alpha|\right), 
\end{equation}
which is a {\it Schr\"odinger cat state\/} rotated $\pi/2$ relatively to the initial
states. This is an unexpected organization, because according to what it is normally 
found in the literature, the field returns to its initial state at most, which is a 
mixed one in this case. 
 
In order to illustrate this peculiar behaviour, we can follow the evolution of the 
field purity, defined as 
\begin{equation}
\zeta_f(t)=1-Tr_{f}\left[ \widehat{\rho }_{f}^{2}(t)\right].
\end{equation}
For an initial statistical mixture we have that
\begin{equation}
\zeta_f(t)=1-\left(
T_{1;+}^{2}+T_{1;-}^{2}+T_{2;+}^{2}+T_{2;-}^{2}
+2\left| T_{3;+}\right| ^{2}+2\left| T_{3;-}\right| ^{2}\right) ,
\label{eq.puresm}
\end{equation}
where

\begin{eqnarray}
T_{1;\pm} &=&\sum_{n=0}^{\infty }P_{n}^{M}\left(\pm 1\right)^{n}C_{n+1}^{2}, \ \ \ \
T_{2;\pm}=\sum_{n=0}^{\infty}P_{n}^{M}\left(\pm 1\right)^{n+1}\left(\frac{n}
{\alpha^2}S_{n}^2\right), \nonumber \\  
T_{3;\pm}&=&\sum_{n=0}^{\infty }P_{n}^{M}\left(\pm 1\right) ^{n+1}C_{n+1} 
\left(\frac{i\sqrt{n}}{\alpha }S_{n}\right),  
\end{eqnarray}
and $P_n^M$ is the Poisson distribution
$P_n^M=e^{-|\alpha|^2}|\alpha|^{2n}/n!$.

As we see in Fig. 1, because the field is 
initially in a mixed state, $\zeta(0)=0.5$. As time goes on we note a growth in
$\zeta$, followed by a sudden decrease, almost down to zero at half of the revival
time. Of course the total atom-field state can not have its purity diminished, 
which means that as the field becomes more pure the atomic state must be 
closer to a mixed state. Although this behaviour is not obvious, 
exists a neat explanation from the phase space point of view, as we are
going to show below. 

\section{Phase space approach}

The representation of fields in phase space has been providing new insights of the
Jaynes-Cummings field dynamics \cite{vidi,risken,matsuo}. Perhaps the most 
convenient quasiprobability to be used in this kind of problem is the $Q$-function, 
defined as
\begin{equation}
Q(x,y;t)=\frac{1}{\pi}\langle\beta|\hat{\rho}_f(t)|\beta\rangle; \ \ \ \ \ \
\beta=x+iy.
\end{equation}
For the specific initial state in Eq.(\ref{eq:insta}), the corresponding 
$Q$-function will be given by
\begin{equation}
Q(x,y;t)=\frac{1}{2\pi }\left( \left| S_{1,+}\right| ^{2}+\left| S_{1,-}\right|
^{2}+\left| S_{2,+}\right| ^{2}+\left| S_{2,-}\right| ^{2}\right),
\end{equation}
where the terms $S_{i,\pm}$ are 
\begin{eqnarray}
S_{1,\pm} &=&\left\langle \beta \right| \widehat{A}\left| \pm\alpha \right\rangle
=\sum_{n=0}^{\infty }\frac{\exp \left[ -\left( \left| \beta \right|
^{2}+\alpha ^{2}\right) /2\right] }{n!}\left( \beta ^{*}\alpha \right)
^{n}\left( \pm 1\right) ^{n} C_{n+1}(t) \\
S_{2,\pm} &=&\left\langle \beta \right| \widehat{B}\left| \pm\alpha \right\rangle
=\sum_{n=0}^{\infty }\frac{\exp \left[ -\left( \left| \beta \right|
^{2}+\alpha ^{2}\right) /2\right] }{n!}\left( \beta ^{*}\alpha \right)
^{n}\left( \pm 1\right) ^{n}\left[-\frac{i\beta ^{*}S_{n+1}(t)}{\sqrt{n+1}}\right]. 
\nonumber
\end{eqnarray}

The $Q$-function shows a very clear picture of the field dynamics. It is already
well-known that for an initial coherent state the collapse is associated to a
split of the $Q$-function in two branches, and that at half of the revival time, 
when the field becomes very close to a pure state, the two branches are most far 
apart \cite{vidi,risken}. In the case of an initial statistical mixture as in 
Eq.(\ref{eq:insta}) (Fig. 2a), there will be counter-propagating branches 
(Fig. 2b), which ``collide'' at half of the revival 
time $(t=\pi/2\lambda)$, as it is illustrated in Fig. 2c. Because we start with an
statistical mixture, this means that we have either one possibility or the other.
It happens that exactly at half of the revival time, there is a complete overlap of
the $Q$-functions representing both possibilities, and also at this time the field
will be in a pure state for each one of them. Because of that overlap, there is 
only one possible state, which happens to be a pure state (Schr\"odinger cat). 

\section{Transition from even to odd-coherent states}

It is worth analysing how would be the dynamics like if the initial field was a 
Schr\"odinger cat state, or
\begin{equation}
\hat{\rho}_f(0)=|\Phi\rangle\langle\Phi|; \ \ \ \ \ \ 
|\Phi\rangle= {\cal N}^{1/2} \left(|\alpha\rangle +r|-\alpha\rangle\right),
\label{eq:statap}
\end{equation}
being $r=\pm 1$ (even and odd-coherent state, respectively).

In this case, the time evolution will be such that
\begin{eqnarray}
\hat{A}|\Phi\rangle&=&{\cal N}^{1/2} \sum_{n}\frac{e^{-|\alpha|^2/2}}{\sqrt{n!}}\alpha^n
\left[[1+r(-1)^n\right]\cos\left[\lambda t(n+1)\right]|n\rangle,\nonumber \\
\hat{B}|\Phi\rangle&=&{\cal N}^{1/2} \sum_{n}\frac{\sqrt{n}}{\alpha}\frac{e^{-|\alpha|^2/2}}
{\sqrt{n!}}\alpha^n\left[[1-r(-1)^n\right]\sin\left[\lambda tn\right]|n\rangle.
\end{eqnarray}
The highly oscillating photon number distribution of the even (odd) coherent state,
\begin{equation}
P_{n}^{S}=\frac{\exp \left( -|\alpha|^{2}\right)|\alpha|^{2n}\left[
1+r(-1)^{n}\right] ^{2}}{n!\left[ 1+r^{2}+2r\exp \left( -2|\alpha|^{2}\right)
\right] },\label{eq:distss}
\end{equation}
is nonzero only at even (odd) photon numbers, and therefore
the first revival with this initial field occurs at half of the time ($t_r^{even}$)
than for a coherent state ($t_r^{cohe}$) \cite{vidi,gerry}, or
\begin{equation}
t_{r}^{even}\left[ 2\lambda t\left( \overline{n}+2\right) -2\lambda t\overline{n%
}\right]\approx 2\pi \ \ \ \ \ \ t_{r}^{even}=\frac{t_{r}^{cohe}}{2}=\frac{\pi}{2\lambda}.
\end{equation}
The expressions in Eq.(\ref{eq:statap}) become, at the revival time 
$t_r=\pi/2\lambda$ and in the limit of $\overline{n}\gg 1$,
\begin{eqnarray}
\hat{A}|\Phi\rangle&=&\frac{i{\cal N}^{1/2}}{2}(1-r)
\left(|i\alpha\rangle-|-i\alpha\rangle\right)\nonumber \\
\hat{B}|\Phi\rangle&=&\frac{{\cal N}^{1/2}}{2i}(1+r)
\left(|i\alpha\rangle-|-i\alpha\rangle\right).\label{eq:statasi}
\end{eqnarray}
We see that for an initial {\it even coherent state\/} $r=1$, 
the field state will become
\begin{equation}
|\psi\rangle=\frac{1}{\sqrt{2}}\left(|i\alpha\rangle-|-i\alpha\rangle\right),
\end{equation}
i.e., a kind of {\it odd-coherent state\/}. However, at the second revival time 
$(t=\pi/\lambda)$, the field will return to its initial state (even-coherent state).
There will be then a periodic change between odd and even coherent states of the
field. Because those states differ by one photon, we expect the atom also to change
state, i.e., to be found in the ground state when the field is in an odd-coherent 
state. 
This is confirmed if we follow the time-evolution of the (periodic) atomic 
population inversion $W(t)=<\sigma_z>$, that can be written in a closed form as 
follows
\begin{eqnarray}
W(t)&=&\left[ 1+r^{2}+2r\exp \left( -2|\alpha|^2\right) \right]
^{\left( -1/2\right)}\{\left( 1+r^{2}\right) \exp
[-2|\alpha|^{2}\sin ^{2}(\lambda t)]\cos [|\alpha|^2\sin (2\lambda
t)+2\lambda t]\nonumber \\  
&+&2r\exp [-2|\alpha|^2\cos {}^{2}(\lambda t)]\cos [|\alpha|^2
\sin (2\lambda t)-2\lambda t]\}. \label{eq.invs}
\end{eqnarray}
In Fig. 3 we have a plot of the atomic inversion as a function of $\lambda t$,
where we see the flip from the upper to lower state occurring periodically.
On the other hand, if the field is initially prepared in an {\it odd-coherent 
state\/} $(r=-1)$, there is never a transformation onto an {\it even-coherent 
state\/}; the field returning periodically to its initial state, as we easily see
from Eq.(\ref{eq:statasi}).

\section{Conclusions}

We have shown that the dynamics of the intensity-dependent Jaynes-Cummings model
makes possible, for sufficient strong fields, to transform a statistical mixture of
two coherent states into a Schr\"odinger cat state. This is a consequence 
of the intrinsic periodicity of the model, and can be readily explained from the
phase-space point of view. We may ask why this behaviour has not been noticed by 
considering the field evolution in the ordinary Jaynes-Cummings model. The answer 
is that despite of the fact that the overlap (at half of the revival time) 
in phase space somehow occurs also in that case, the precise 
match between the clockwise and the counter-clockwise branches hardly happens. 
This is due to deformations of the branches as time goes on, and the less than
perfect overlap means that we continue having two possible (mixed) states, although
there is a tendency of purification. Nevertheless, perfect purification is not achieved
in this case.

Regarding the transformation of the field from an even-coherent state to an 
odd-coherent state, the atom must change its state not only for energy 
conservation reasons,
but also for parity conservation in the atom-field system. Both atom and field
have well defined parity, and the change of atomic parity due to the transition
from the excited state to the ground state is compensated by the change of the
field from the even-coherent state to the (rotated $\pi/2$) odd-coherent state.

\newpage 

\begin{figure}
\vspace{0.5cm}
\centerline{\psfig{figure=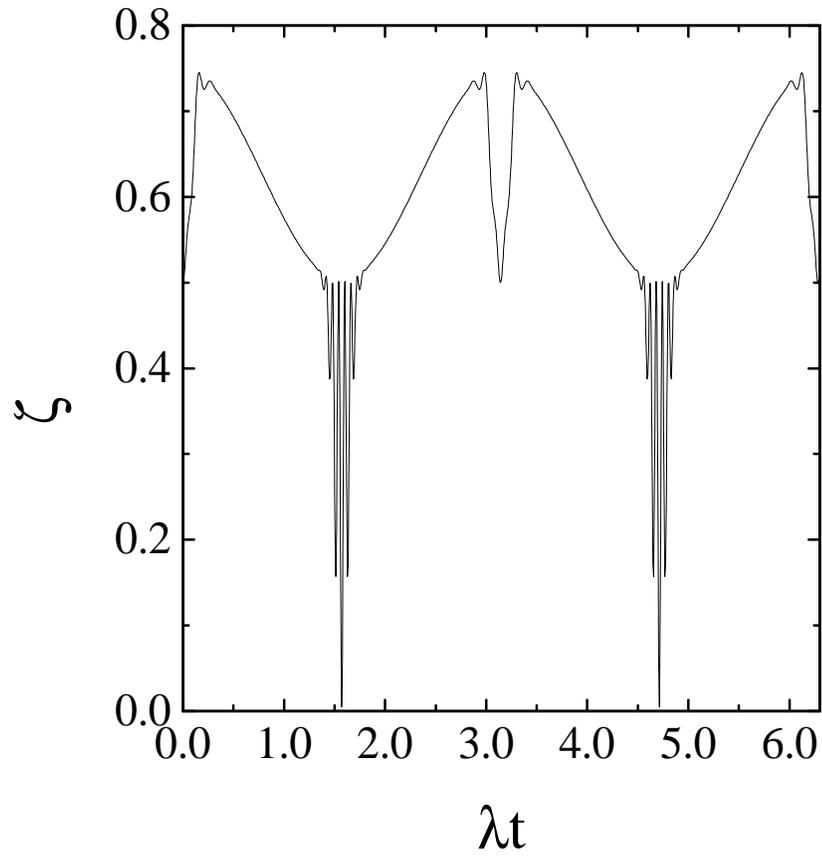,height=18.0cm}}
\vspace{1.2cm} 
\caption{Field purity parameter $\zeta$ as a function of time for an initial
statistical mixture of two coherent states 
$|\alpha\rangle$ and $|-\alpha\rangle$ ($\alpha=5$).}
\end{figure}

\newpage

\begin{figure}
\vspace{0.5cm}
\centerline{\psfig{figure=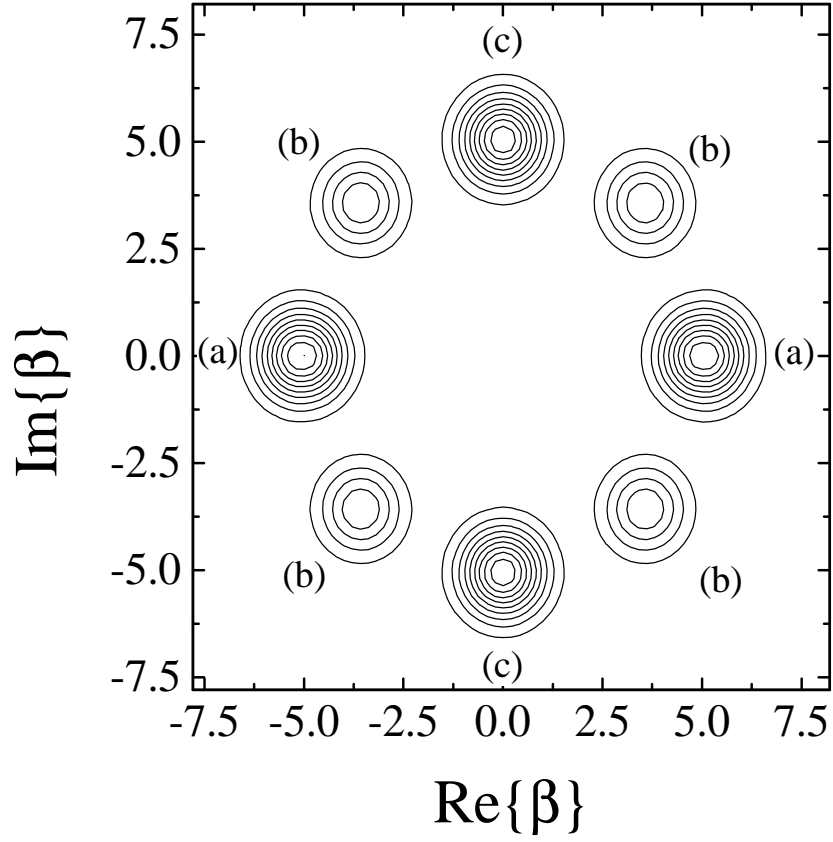,height=18.0cm}}
\vspace{1.2cm} 
\caption{$Q$-function for an initial statistical mixture of two coherent states
$|\alpha\rangle$ and $|-\alpha\rangle$ ($\alpha=5$), at different times, (a) $t=0$,
(b) $t=\pi/4\lambda$ and (c) $t=\pi/2\lambda$.}
\end{figure}

\newpage

\begin{figure}
\vspace{0.5cm}
\centerline{\psfig{figure=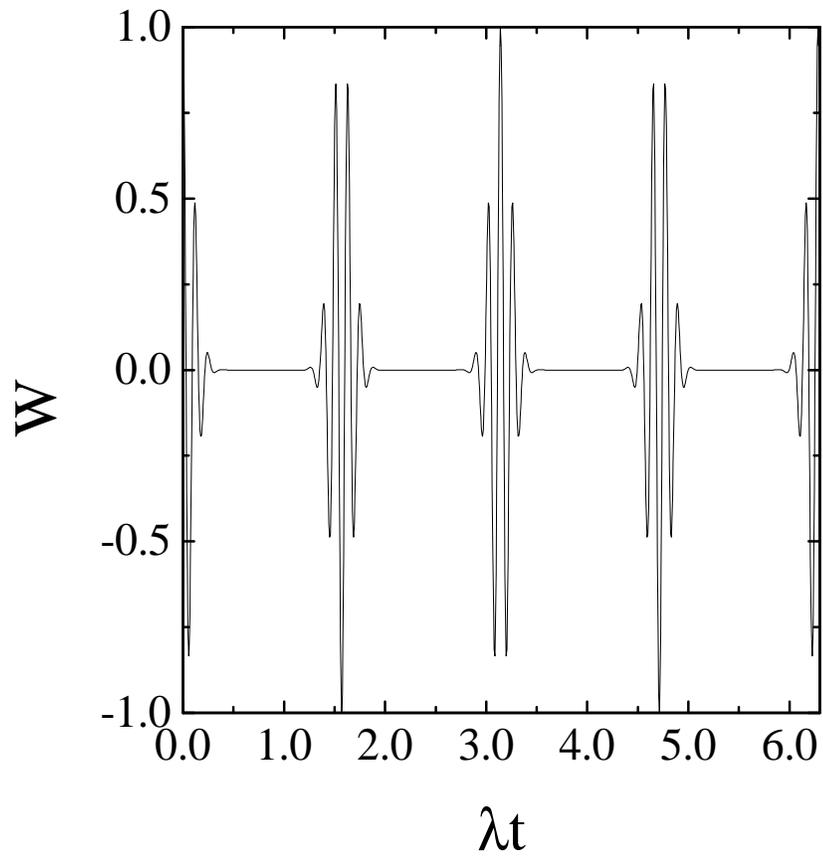,height=18.0cm}}
\vspace{1.2cm} 
\caption{Atomic population inversion for the field initially prepared in an 
even-coherent state and the atom in the excited state with $\overline{n}=25$.}
\end{figure}

\acknowledgements{We would like to thank Mr. D. Jonathan for useful comments.
This work was partially supported by CAPES$^{*}$ (Coordena\c c\~ao de Aperfei\c
coamento de Pessoal de N\'\i vel Superior, Brazil), and CNPq$^{\S}$ (Conselho
Nacional para o Desenvolvimento Cient\'\i fico e Tecnol\'ogico, Brazil).}

\end{document}